\documentstyle[12pt]{article}

\input epsf
\epsfverbosetrue

\global\arraycolsep=2pt
 \newcommand{\be}{\begin{eqnarray}}
\newcommand{\ee}{\end{eqnarray}}
\newcommand{\la}{\langle}
\newcommand{\ra}{\rangle}
\begin{document}
  \begin{titlepage}
\begin{center}

{\Large \bf  Condensates and Vacuum Structure of Adjoint $QCD_2$}
 \vspace{2cm}

  {\bf T. Fugleberg}, {\bf I. Halperin} and {\bf   A. Zhitnitsky }
 \vspace{0.5cm}

 {\it Department of Physics and Astronomy, University
 of British Columbia, 6224 
Agricultural Road, Vancouver, BC, V6T 1Z1,  Canada}
\date{\today}
\end{center}
\begin{abstract}
We discuss a  two dimensional $SU(N)$ 
Yang-Mills theory coupled to  massive  adjoint fermions
for different worlds classified by the integer $k=0,1,...N-1$. 
We study the fermion condensate for these unconnected worlds 
as a function of the parameter $k$. We show that the condensate as well 
as the spectrum of the theory do depend on this vacuum parameter $k$.
  
Technically, the value of the fermion condensate is related 
to the value of the gluon condensate via the Operator Product Expansion.
We use this to find the leading dependence (in the limit of a heavy quark) 
of the fermion condensate $\la  k |\bar{\psi} \psi| k \ra $
on the nontrivial vacuum angle $k$.  We also determine the gluon condensate 
of the theory using low-energy theorems.
\end{abstract}
\end{titlepage}
\vskip 0.3cm
\noindent

\section{Introduction}
Two dimensional $SU(N)$ Yang Mills theories coupled to adjoint fermions
continue to attract theorists \cite{Klebanov}-\cite{Christiansen} in view
of their interesting connections with both string theory on one side and 
real four dimensional $ QCD $ on the other. In particular,
as in the latter, $QCD_2 $ with adjoint matter is 
expected to possess a nonvanishing fermion condensate
$\la\bar{\psi}\psi\ra $. 
This corresponds to a nontrivial vacuum structure of the theory.
It is also believed that such a structure is a consequence of nontrivial
topological properties of the group $SU(N)/Z_N$, which is 
the relevant
symmetry group when adjoint matter (rather than fundamental matter)
is considered\footnote{see however, Ref.\cite{Christiansen} where 
it is argued
that these topological properties are relevant for a theory with
fundamental matter also.}. Therefore, it is expected that the theory 
possesses a nontrivial vacuum angle $k$ \cite{Witten}, similar to the
$\theta$ angle of QCD. It is known that many problems such as 
the $U(1)$ problem, chiral symmetry breaking phenomenon, confinement
and multiplicity of vacuum states are related to each other and to
$\theta$ dependence. Therefore, in spite of the fact that we live in 
a certain vacuum state  (this is expressed 
by the so-called superselection rule),
the physical parameters of the theory do depend on $\theta$, which is an
extremely important characteristic of the theory.

The main goal of the present paper is an analysis of the 
condensate $\la k| \bar{\psi}\psi|k\ra $ as 
a function of the discrete topological vacuum angle 
label $k=0, 1, ... N-1$.  This dependence 
can be found explicitly as we shall see.
Therefore, the vacuum condensate could be considered as an order parameter
which distinguishes different vacuum states.

First we would like to give a short historical introduction
of the subject.
For the case of fundamental fermions in the large N limit 
('t~Hooft model \cite{'t Hooft})
and vanishing fermion mass, the fermion condensate was first calculated in 
\cite{Zhitnitsky}. Later on, the quark condensate was calculated 
in several different ways, both analytically and numerically 
\cite{Li}\cite{Lenz}\cite{Burkardt2}. 

In the case of adjoint fermions, using 
bosonization    it has been argued
  \cite{Smilga} that the
condensate arises for arbitrary N.  At the same time,
instanton calculations seem to imply the vanishing of the 
condensate in the chiral limit   for $N\geq 3$.
 An independent argument, based on quark-hadron duality, has been 
given to support a nonzero value for the condensate 
in the large N limit\cite{Kogan_Zhitnitsky}.
On the formal side,  in the small volume limit the condensate 
was calculated
for $SU(2)$ and $SU(3)$ gauge groups using ET quantization 
\cite{Lenz_etal}.
Similar calculations have been performed for the $SU(2)$ theory in
finite volume $L$
using light cone quantization (see  
    \cite{Pinsky_Mohr} \cite{McCartor_etal}).  In all 
finite volume calculations
 the value of the condensate appears to scale as 
$\frac{1}{L}$, where L is the quantization volume. Therefore, no 
continuum limit can be taken in these formulae,
and no finite result can be obtained. However, 
 it is believed that 
an extra factor of $L$ will appear in a complete theory 
leading to a finite value for the condensate as in the 
Schwinger model \cite{McCartor_etal} \cite{McCartor}.  
Unfortunately, this would require a complete solution of the 
problem which is not possible at present and thus the 
question regarding the magnitude of the condensate in the 
continuum limit can not be answered within this approach 
at the moment. 

We conclude this short historical introduction by emphasizing that 
no meaningful calculation for the quark condensate is hitherto available
in the continuum limit for two dimensional $QCD$ with adjoint matter. 
Therefore, discussion of the number 
of different  vacuum states    which  
is based on the presumption of a nonzero value for  
the quark condensate has no solid basis before a finite   
value for  the condensate (as an order parameter) is obtained.

Before proceeding we should give a definition of the quark condensate in
the theory.  We define the fermion condensate in  PCAC terms\cite{Adler}
\footnote{ The singlet axial current
in $QCD_2$ is anomaly-free unlike the Schwinger model 
where there is an extra term due to the anomaly.}:
\be 
\label{1}
0&=&\lim_{q \rightarrow 0} i q^{\mu}\int d^2x e^{iqx}\langle 0 |T\{\bar{\psi}\gamma_{\mu}\gamma_5\psi(x) \bar{\psi}i\gamma_5\psi(0)| 0 \rangle \nonumber
\\ &=& 2i\langle 0 |\bar{\psi} \psi| 0 \rangle - 2 m \int d^2x \langle 0 |T\{\bar{\psi}i\gamma_5\psi(x) \bar{\psi}i\gamma_5\psi(0)| 0 \rangle .
\ee
This is the standard nonperturbative definition 
of the quark condensate in terms of 
the original $\psi$ fields. In particular, in the chiral limit 
$m\rightarrow 0$, only a Goldstone boson with 
a finite  (at $m\rightarrow 0$) residue
 $\la 0 | \bar{\psi}i\gamma_5\psi(x)|\pi\ra=
i\frac{f_{\pi}m_{\pi}^2}{2m}$  contributes to the
correlation function $\la 0 |T\{\bar{\psi}i\gamma_5\psi(x) \bar{\psi}i\gamma_5\psi(0)| 0 \rangle$.   
 In this case eq.(\ref{1}) gives the famous relation
$m_{\pi}^2f_{\pi}^2=-4 m\langle 0 |\bar{\psi} \psi| 0 \rangle$
 between the mass of the Goldstone $\pi$ meson  and the chiral condensate.

We want to emphasize that relation (\ref{1}) 
is valid  not only in the chiral limit $m\rightarrow  0$,
but is also satisfied for arbitrary $m$. We expect, of course,
that at large $m\rightarrow\infty$ the condensate goes to zero like 
$\langle 0 |\bar{\psi} \psi| 0 \rangle\sim \frac{1}{m}$.

Nevertheless, 
a small, but not exactly vanishing  condensate will
play its role: {\bf it will be an order parameter which
labels the  different vacuum states}
$|k\ra$ we have been talking about. 
This is the cornerstone of our approach.
We shall calculate the fermion condensate
in the limit of large mass and we shall find different
magnitudes for different $|k\ra$. Presumably 
a calculation at large $m$ corresponds to
the weak coupling regime.  Therefore, a perturbative
calculation is justified. 
At the same time the small non-vanishing condensate
is sensitive to the particular $|k\ra$-vacuum state and 
has a different value for each vacuum state.

As the first step in this program, 
we shall test our conjecture in the 't~Hooft model \cite{'t Hooft}.
This model is well understood in the limit of large $N$.
We know the spectrum as well as 
 all relevant matrix elements. 
Therefore, one can use definition (\ref{1})
in order to calculate the condensate for arbitrary
$m$ exactly. The corresponding  formula is also known and is given
by\cite{Burkardt2}:
\begin{equation}
\label{2}
\langle 0 |\bar{\psi} \psi| 0 \rangle_{ren} = 
\frac{m N}{2\pi} \left[ \ln(\pi\alpha) - 1 - \gamma_E +\left(1-\frac{1}{\alpha}\right)
 [(1-\alpha)I(\alpha) - \ln 4] \right],
\end{equation}
where $\alpha=\frac{g^2C_2(F)}{\pi m^2}$ is the 
dimensionless 
parameter of the theory.
We shall see in section 2 that a
perturbative calculation is justified in the large $m$
limit\footnote{It
corresponds to a 
small parameter $\alpha$ or small coupling constant $g$.} 
in this model.

Encouraged by these results, 
 we go one step further in section 3 and carry out the same 
perturbative calculations for the  
case of adjoint fermions 
with arbitrary $N$\footnote{Note that
this   model   has  not been solved yet,
even in the large N limit.  Therefore, no information
similar to the 't~Hooft case is available.}.
We next use the Operator Product Expansion in the limit 
$m\rightarrow\infty$ in order to relate 
the fermion and gluon condensates.  Nontrivial, 
nonperturbative physics
comes into the game through the gluon condensate.
The dependence of the gluon condensate on the nontrivial 
vacuum structure has been determined
earlier \cite{Paniak_etal} in pure Yang-Mills theory 
in 2 dimensions
by means of the well developed
machinery of Wilson loop calculus. 
 
Finally, in section 4, we use low energy theorems and the
 form of the fermion condensate we obtained to determine 
the $\frac{1}{m^2}$ corrections to 
the gluon condensate due to a large, but finite quark mass $m$.

\section{Lessons from  the 't~Hooft Model}
The 't~Hooft model\cite{'t Hooft} consists of quarks 
interacting via gluons of the SU(N) gauge group with Lagrangian:
\begin{equation}
{\mathcal L}=-\frac{1}{4}G^a_{\mu\nu} G^{a\mu\nu} + 
\bar{\psi}\left( 
i\gamma^{\mu} D_{\mu} - m \right) \psi,
\label{tHooftLagrangian}
\end{equation}
where
\begin{eqnarray}
G^a_{\mu\nu}&=&\partial_\mu A_\nu{}^a - \partial_\nu 
A_\mu{}^a - g f^{abc} A_\mu{}^b A_\nu{}^c, \\
D_{\mu}\psi &=&\partial_{\mu}\psi + i g A_{\mu}{}^a T^a \psi,
\end{eqnarray}
and $T^a$ are the Hermitian generators of the group 
representation.  In the
fundamental representation they are the matrices 
$\lambda^a/2$.
This formulation of the theory is similar to 
\cite{Callan_etal} and differs from the 't~Hooft 
form in that here the gauge group is SU(N) instead
of U(N).  As mentioned in \cite{Callan_etal} the U(N)
singlet decouples and describes a free field and to leading
order in 1/N the distinction does not matter.

In light cone coordinates: 
\begin{equation}
x^{\pm}=\frac{1}{\sqrt{2}}(x^1\pm x^0),
\end{equation}
the problem becomes very simple in the light cone gauge, $A_-=0$.
The Lagrangian becomes:
\begin{equation}
{\mathcal L}= \frac{1}{2} Tr(\partial_-A_+)^2 + \bar{\psi}\left( 
i\gamma^{\mu} D_{\mu} - m \right) \psi.
\end{equation}
There is no ghost in this gauge.  Take $x^+$ as the time 
coordinate and notice
that $A_+$ is not a dynamical field but provides a (non-local) 
Coulomb
force between the fermions.

The algebra for the gamma matrices is:
\begin{eqnarray}
\gamma_-^2=\gamma_+^2=0 \\
\gamma_+\gamma_- + \gamma_-\gamma_+=2.
\end{eqnarray}

\begin{figure}
\epsfysize=2.5in
\epsfbox[-80 505 361 744]{figure1.epsf}
\caption{Feynman rules: where i,j are matrix indices of the 
group representation and a,b are vector indices in the group space.}
\label{feynmanrules}
\end{figure}

The Feynman rules in this case are shown in 
Fig.(\ref{feynmanrules}).

Considering the limit $N\rightarrow\infty$ with $g^2N$ 
fixed corresponds to only
keeping the planar diagrams.
't~Hooft solved for the dressed propagator in this case using 
the Bethe-Salpeter
equation.  The form of the dressed propagator in our notation is:
\begin{equation}
\delta_{ij} \; \frac{i(\gamma_- p_+ + \gamma_+ p_- + 
m)}{2p_+p_- -  m^2 + g^2 N/\pi -
g^2 N |p_-|/\pi\lambda - i\epsilon},
\end{equation}
where $\lambda$ is an infrared cutoff parameter that 
removes the point $k_-$=0
from the momentum space.   't~Hooft found that $\lambda$
drops out in all gauge invariant quantities. This regularization 
has been argued \cite{Callan_etal} \cite{Einhorn} to be equivalent
to a principal value regularization procedure where the photon propagator 
is replaced by it's Cauchy principal value (CPV):
\begin{equation}
\frac{1}{k_-^2}=P \frac{1}{k_-^2}.
\end{equation}
In this form $\lambda^{-1}$ is actually a gauge parameter
which makes its cancellation from gauge invariant quantities obvious.
With this regularization prescription the propagator is identical to the
't~Hooft propagator except we neglect the terms involving $\lambda$.
 
With the motivation presented in the Introduction,
we now want to calculate the value of the condensate
in the well understood 't~Hooft model 
in the large $m$ limit. We achieve this by two independent
approaches.  First, we do the standard perturbative calculations which are 
perfectly justified in the  large $m$ limit. The second way
of doing the same physics is to make use of the dressed 
't~Hooft propagator which is a solution of the Bethe-Salpeter equation.
Agreement of the obtained results should be 
considered   as a test of the method where a nonperturbative 
condensate (\ref{1}) in the large $m$ limit is simply determined by a
pure perturbative calculation.

\begin{figure}
\epsfysize=2.25in
\epsfbox[70 442 609 667]{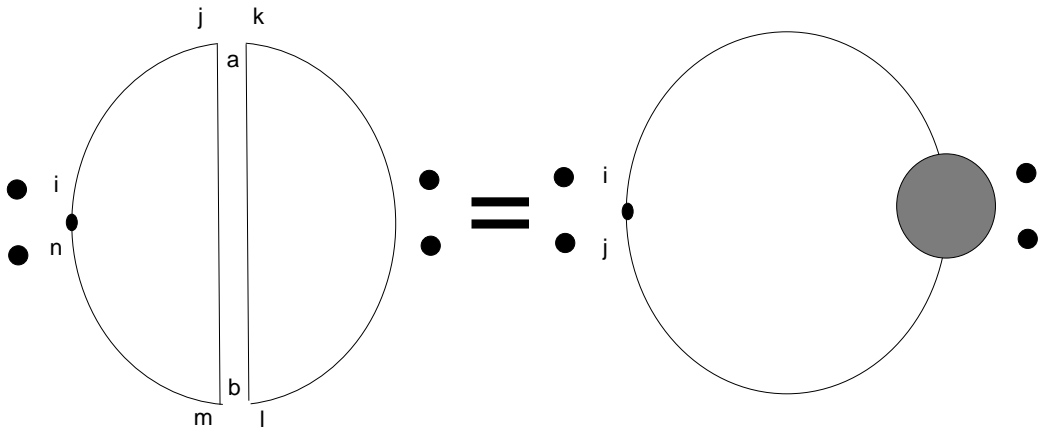} 
\caption{Bare and dressed vacuum loop calculations shown to agree to order $g^2$.}
\label{loopdiagrams}
\end{figure}

The first approach is to calculate $\la\bar{\psi} \psi\ra$
from the first diagram in Fig.(\ref{loopdiagrams}).
We calculate this bare diagram using both the 't~Hooft
regularization scheme and principal value regularization and 
found that
they are equivalent at least to order $g^2$:
\begin{equation}
\label{3}
\la\bar{\psi} \psi\ra  = - N \;  C_2(F) \; \frac{1}{\pi^2} 
\frac{g^2}{m} = 
 - \frac{(N^2-1)}{2} \;  
\frac{1}{\pi^2} \frac{g^2}{m},
\end{equation}
where the second Casimir constant of the fundamental representation 
arises
from $T^a_{ij} T^a_{ji}= Tr(C_2(F) {\mathbf I}_F)= C_2(F) N$ after 
contraction over group matrix
indices because of the closed fermion loop.
We note that in the large $N$ limit where
$g^2N\sim const.$, the condensate $\la\bar{\psi} \psi\ra  \sim N $
as it should. We also note that in the large $m$ limit it goes 
to zero as $1/m$
in agreement with the general discussions presented in the 
Introduction.

The second approach to the same 
calculation is to take the dressed 't~Hooft propagator
and expand it to
the first order in $g^2$
(from the definition
of the condensate we have to take the difference between the
 dressed and free propagator):
\begin{eqnarray}
\int \frac{dp_- dp_+}{(2\pi)^2} \left[  \right. &\delta{ij}& \;  \frac{i(\gamma_- p_+ + \gamma_+ p_- + m)}{2p_+p_- -  m^2 + g^2 N/\pi -
g^2 N |p_-|/\pi\lambda - i\epsilon} \\
  &-\delta{ij}& \;\frac{i(\gamma_- p_+ + \gamma_+ p_- + 
m)}{2p_+p_- - m^2 -i\epsilon} \left. \right]. \nonumber
\end{eqnarray}
For small values of g we can expand the denominator of the first term
so that we have:
\begin{eqnarray}
&&\frac{1}{2p_+p_- -  m^2 + g^2 N/\pi - g^2 N |p_-|/\pi\lambda - 
i\epsilon} = \nonumber \\
&&\frac{1}{2p_+p_- -  m^2 - g^2 N |p_-|/\pi\lambda - i\epsilon} 
\left[ 1 - \frac{g^2N/\pi}{2p_+p_- -  m^2 -
 g^2 N |p_-|/\pi\lambda - i\epsilon} \right].
\nonumber
\end{eqnarray}
The first term of the expansion cancels the bare propagator in CPV
regularization where terms containing $\lambda$ are simply dropped.
Therefore we have the integral:
\begin{eqnarray}
\int \frac{dp_- dp_+}{(2\pi)^2}  \; \delta{ij} 
\frac{g^2N/\pi}{[2p_+p_- -  m^2
  - i\epsilon]^2}.
\end{eqnarray}

The result is the same as above (\ref{3}) in the large $N$ limit
 and agrees with the large m 
expansion of  formula (\ref{2}), (see  Ref.\cite{Burkardt2}).  The moral of
this section is that
 we are confident that the
perturbative calculation gives the
correct expression for the condensate
(\ref{1}) in this completely solvable example.  
Now we would like to apply this approach
to a model which (unlike the 't~Hooft model) is not completely 
solvable,
but is much more interesting because it includes nontrivial
vacuum structure labeled by an integer number $k=0, 1, ...N-1$.
We consider this problem in the next section.

\section{Condensate with Adjoint Fermions}

Now we want to do the same calculation for two dimensional QCD where
 we take the fermions to be in the adjoint representation.  The  
QCD lagrangian is the same as before 
with the only difference that the generators of the group  are  
the structure
constants of the group
\begin{equation}
(T^a)_{bc}= - i f_{abc} ~~~~~~~~~ a=1,2 \ldots (N^2-1).
\end{equation}
 
We now repeat our previous calculation
 of the first diagram in Figure (\ref{loopdiagrams}) 
with the following result:
\begin{equation}
\la\bar{\psi} \psi\ra  = - (N^2-1)\;  C_2(A) \; 
\frac{m}{\pi^2} \frac{g^2}{m^2} = 
- N(N^2-1)  \;  \frac{g^2}{\pi^2 m},
\label{adjointcondensate}
\end{equation}
where we take into account the value for the second Casimir constant
for the adjoint representation:
\begin{equation}
T^a_{ij} T^a_{ji}= Tr(C_2(A) {\mathbf I}_A)= C_2(A) (N^2-1).
\end{equation}
This result corresponds to the trivial vacuum, i.e. $k=0$.
 We would like to do the same perturbative calculations for a 
nontrivial vacuum 
state $|k\ra$ as well.  Recall that 
the case of fermions in the adjoint representation has long been known
\cite{Witten} to possess $N$ different vacua corresponding 
to fundamental  
external charges at the
 boundaries of the universe.  This is equivalent to a
universe with a background color electric field analagous to the
$\theta$-vacua in Coleman's analysis of the  
Schwinger model\cite{Coleman}.
Different  vacua in the nonabelian case have, in general, 
different vacuum energy and can be labeled by discrete
values of $\theta$:
\begin{equation}
\theta_k=\frac{2\pi}{N} k \; \; \; k=0,1, \ldots (N-1).
\end{equation}
These vacua are unstable when the dynamical fermions are in 
the fundamental
representation because they can screen the charges at the edges of
the universe to form a trivial vacuum\cite{Witten}.  This is the 
analogy 
of pair creation in the massive Schwinger model.  When the fermions 
are in the adjoint
 representation such a  screening
 cannot occur and therefore there is a nontrivial vacuum structure
and consequently  a  nontrivial dependence of the  observables
on this vacuum angle $k$.

Now the question is: {\bf Is it possible
to obtain a nontrivial dependence on the topological angle 
$k$ from the perturbative calculation},  
which apparently does not contain any information regarding  
the topological properties
of the theory at all?  Are we able to see the effect of the 
vacuum angle $\theta_k$ in a perturbative expansion
$\sim \sum_nc_n(\frac{g^2}{m^2})^n$ or is this impossible in principle?

\begin{figure}
\epsfysize=2in
\epsfbox[-140 422 331 686]{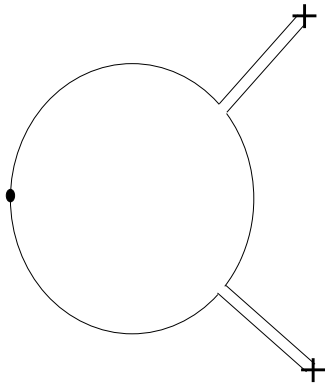}
\caption{Heavy fermion loop in an external gluon field}
\label{externaldiagram}
\end{figure}

The answer to the question formulated above is: yes, we can 
extract information
about vacuum topological properties of the theory.  In fact, this
information is contained in the coefficients $ c_n $ of the 
perturbative expansion in powers of $ g^2/m^2 $. The point
is that external charges at the boundaries of the universe
cannot, of course, talk in the gauge theory to vacuum 
fermions directly, but they can communicate via gluons
as soon as the external charges act as sources for the 
latter. It is therefore clear that in order to see a 
nontrivial $k-$dependence in the fermion condensate,
$ \la \bar{\psi} \psi \ra $, we should go beyond the 
leading term (\ref{adjointcondensate}) in the expansion
over $ g^2 / m^2 $ and take into account terms of higher 
orders in $ 1/m $ which are accompanied by powers of the
gluon field. Technically, this problem amounts to a 
calculation of the heavy fermion loop in an external
gluon field (see Fig.(\ref{externaldiagram})) which does 
depend on the topological number $ k $. In our case 
both the use of the perturbative fermion propagator and 
retaining only a finite number of powers of the 
external field (and its derivatives) are justified by the 
smallness of the parameter $ g^2 / m^2 $. This procedure
results in a particular form of the Operator Product
Expansion in powers of $ 1/m $ (known also as the heavy
quark expansion). Note that by gauge invariance 
the expansion in external field starts from a term of 
second order in the field,   
and this is why we expect 
to get a nonzero effect at the level of order $g^4/m^3$.

Therefore, at large $m$ we are justified in the 
use of perturbation theory in order to reduce
the problem of the calculation of $\la k| \bar{\psi} \psi |k\ra$
to the problem of the calculation of the gluon condensate
$\la k| Tr G_{\mu\nu}  G^{\mu\nu}|k \ra $.
The corresponding method is well-known
and we refer to a nice technical review\cite{NSVZ}
of the subject.
The result of the calculation is:\footnote{We note that the 
analogous formula in real four dimensional
$QCD_4$ takes the form\cite{NSVZ}:
$\la  \bar{\psi} \psi \ra 
 = -\frac{g^2}{48\pi^2m} 
  \la    G_{\mu\nu}^a G^{a\mu\nu}\ra $.}:
\begin{eqnarray}
\la k | \bar{\psi} \psi | k \ra 
 -\la 0 | \bar{\psi} \psi | 0 \ra  
=&-&\frac{1}{12 \pi} \frac{g^2 N}{ m^3} \left( \la     
k | G_{\mu\nu}^a  G^{a\mu\nu} | k \ra
 - \la 0 |     
G_{\mu\nu}^a  G^{a\mu\nu}| 0 \ra  \right) \nonumber  \\ &+&O(1/m^5).
\label{OPE}
\end{eqnarray}
The problem of the dependence of the gluon condensate $\la k |    
G_{\mu\nu}^a  G^{a\mu\nu}| k \ra$ on nontrivial vacuum structure 
in pure gluodynamics in two dimensions has 
been solved earlier in \cite{Paniak_etal}. The dynamical 
heavy quarks with mass $m$ will bring some order $1/m^2$ corrections 
to this result and we neglect these corrections 
at the moment (see the corresponding calculations in the next section).
We quote the relevant formula for the gluon condensate  in 
Minkowski space
rather than in 
Euclidean space, where it was originally
obtained \cite{Paniak_etal}: 
\begin{equation}
\label{G}
\la k|   G_{\mu\nu}^a   G^{a\mu\nu}|k \ra 
-\la k=0|   G_{\mu\nu}^a   G^{a\mu\nu}|k =0\ra   = g^2 
\frac{2k(N-k)(N+1)}{N}.
\end{equation}
Therefore, the final expression for the condensate as 
a function of 
the vacuum label $k$ takes the following form:
\begin{equation}
\la k|\bar{\psi} \psi |k\ra 
- \la 0|\bar{\psi} \psi |0\ra = 
-\frac{N}{6\pi} \frac{g^4}{m^3}  \frac{k(N-k)(N+1)}{N}.
\label{OPE1}
\end{equation}
 
We close this section with a few remarks.  First,  
as   was expected, a nontrivial dependence on $k$ 
appears in the final formula only at the level $\sim g^4/m^3$
and therefore is relatively small in large $m$ limit. 
Nevertheless, formula (\ref{OPE1}) 
demonstrates that all physical properties are different for
 different 
vacua label $k$.  Vacuum condensates are different.  String
tensions\cite{Paniak_etal} are different and therefore the mass
spectrum is different.  Vacuum energies are also different.
We demonstrated this fact by keeping only the leading $1/m$ 
terms.
However, we believe that the statement has a more
general origin and it is hard to believe that different 
vacua could become
identical for 
a finite value of $m$
as long as the two parameters $ m $ and 
$ k $ are independent of each other. Our last remark 
regarding formula 
(\ref{OPE}) is that the function which 
appears there has   
a property which was 
expected from the very beginning\cite{Witten} - it is a periodic
function such that 
states $|k=0\ra$ and $|k=N\ra$ are equivalent. Therefore,
we do not obtain a new state each time we increase the label $k$.
Instead, we return to the starting state after $N$ steps 
have been made. 
We notice also the symmetry property $k\Leftrightarrow (N-k)$
which was also expected\cite{Paniak_etal}.

\section{Low Energy Theorems}
The purpose of this section is twofold. First of all, the 
low energy theorems
make it possible to investigate the behavior  of vacuum 
condensates
with changing quark mass $m$ in the vicinity of parameters 
where
those condensates are known, i.e. at $m\rightarrow\infty$.
This information might help  in  understanding the qualitative 
features of the model in general and its vacuum structure 
in particular.
The second goal of this study is the calculation of the 
$\frac{1}{m^2}$ corrections 
to the gluon condensate 
calculated in pure gluodynamics \cite{Paniak_etal} due to 
the presence of 
matter fields. 

A similar study  of the low energy theorems
in the 't~Hooft model was carried out  for the first 
time in ref. 
\cite{Zhitnitsky}, for vanishing fermion mass, where 
the corresponding
quark and gluon condensates have been calculated. This 
program was pushed forward
in Ref.\cite{Burkardt2} where the previous result was 
generalised to
arbitrary quark mass in the 't~Hooft model.  

The main idea of the derivation of the low energy theorems 
is quite simple\cite{Novikov}.
From the definition of the Path Integral, the variation of 
the gluon condensate
with mass $m$ is determined by the following correlation function
\begin{equation}
\label{4}
\frac{d}{dm}\la G^{a}_{\mu\nu} G^{a\mu\nu} \ra=
-i\lim_{q \rightarrow 0}
\int d^2x e^{iqx}\la 0|T\{\bar{\psi} \psi (x),   G^{a}_{\mu\nu} G^{a\mu\nu}(0) \} |0\ra . 
 \end{equation}
At the same time, the obtained correlation function can be 
explicitly calculated due to the following identity\footnote{
To derive this identity we simply rescale the gluon 
field $A_{\mu}^{'}=
gA_{\mu}$ such that dependence on
$g$ appears 
in the Lagrangian 
only in the combination $-\frac{1}{4g^2}G^{a'}_{\mu\nu} 
G^{a'\mu\nu}$.}
\begin{equation}
\label{5}
i\lim_{q \rightarrow 0}
\int d^2x e^{iqx}\la 0|T\{\bar{\psi} \psi (x),   G^{a}_{\mu\nu} G^{a\mu\nu}(0) \} |0\ra = 
-\frac{4}{g^2}\frac{d}{d(1/g^2)}\la \bar{\psi} \psi\ra .
\end{equation}
Up to this point the derivation in $QCD_2$ is identical to 
that in $QCD_4$. 
The only relevant point is the form of the Lagrangian 
(\ref{tHooftLagrangian})
and not the specific properties of the
theory which of course depend on the dimensionality of 
the space-time. The difference appears in the explicit 
calculation of the right hand side of Eq.(\ref{5}). 

In   $QCD_4$ the corresponding calculation
\cite{Novikov} of $ \frac{d}{d(1/g^2)}\la 
\bar{\psi} \psi\ra$
is based on the observation that due to the asymptotic freedom 
and to the absence of mass parameter other than 
$M$ (the ultraviolet cut-off),
the   dependence
on $g$ comes exclusively through 
$\Lambda_{QCD}\sim M\exp{-\frac{8\pi^2}{bg^2(M)}}$ 
where $b=11N/3-2N_f/3$
is the first term in the Gell-Mann-Low $\beta$-function.

In the present case of $QCD_2$ the dependence on $g^2$ is 
known exactly in 
large $m$ limit (\ref{OPE1}). Therefore, we can 
explicitly calculate the behavior  of the vacuum condensate 
$\la G^{a}_{\mu\nu} G^{a\mu\nu} \ra$ 
 with variation of the  quark mass $m$:
 \begin{equation}
\label{6}
\frac{d}{dm}\left(\la G^{a}_{\mu\nu} G^{a\mu\nu} \ra_k-
\la G^{a}_{\mu\nu} G^{a\mu\nu} \ra_0\right)=
 \frac{N}{\pi} \frac{4g^4}{3 m^3}  \frac{k(N-k)(N+1)}{N}. 
 \end{equation}
Note that in this formula we keep only the information 
relevant to the $k$-vacuum  dependence.  As mentioned above,
there is a leading term ($\sim g^2/m$) in the fermion 
condensate (\ref{adjointcondensate})
which does not contain any $k-$dependence.   
A similar term in the gluon condensate depends on 
subtraction and normalization procedures but we will 
not discuss this issue here. However, the difference 
in condensates between two vacuum states
has an absolute meaning and the corresponding expression 
is given by eq. (\ref{6}).

Formula (\ref{6}) makes it possible to calculate the finite 
correction due to the dynamical heavy quark to the gluon 
condensate calculated previously \cite{Paniak_etal} 
in pure gluodynamics. Integrating Eq.(\ref{6}) at large $ m $,
we obtain
\begin{equation}
\label{7}
\la k|   G_{\mu\nu}^a   G^{a\mu\nu}|k \ra 
-\la  0|   G_{\mu\nu}^a   G^{a\mu\nu}| 0\ra   = 
g^2 \frac{2k(N-k)(N+1)}{N}
(1-\frac{1}{3}\frac{g^2N}{\pi m^2}).
\end{equation}
The correction $\sim \frac{g^2N}{\pi m^2}$ in this formula 
is not suppressed
in the large $N$ limit.  This is in accordance with the 
general expectation
that fluctuations of adjoint matter fields (unlike 
fundamental matter)
are not suppressed by a factor $1/N$.

One can go a little bit further in the analysis of the 
low energy theorems 
by taking advantage of 
the specific property of two dimensional $QCD$ that
the only dimensionless combination in the theory is 
$g^2/m^2$.   Therefore, without loss of generality
one can present the fermion condensate  in the following
  form:
\begin{equation}
\la \bar{\psi} \psi\ra = m f(\frac{g^2}{m^2}),
\end{equation}
with some function $f(x=\frac{g^2}{m^2})$.
In this case the derivative with respect to  $m$ and derivative 
with respect to $g$ are related to each other. At the same 
time, as we observed earlier, the derivative  
$\frac{d}{d(1/g^2)}\la \bar{\psi} \psi\ra$ 
is reduced to some  correlation function (\ref{5}) which 
describes the 
dependence of the gluon condensate on $m$ (\ref{4}).
Therefore one can relate the gluon and quark condensate 
exactly without any approximations using only very general
properties of the theory. 
Indeed, from identities
\begin{eqnarray}
\frac{d}{dm}\left( \frac{\la \bar{\psi} 
\psi\ra}{m} \right) 
= -2 \frac{g^2}{m^3} f^{\prime}(x),    \nonumber \\
\frac{d}{d(1/g^2)} \la \bar{\psi} \psi\ra = - 
\frac{g^4}{m} f^{\prime}(x),
\end{eqnarray}
one obtains:
\be
\label{8}
\frac{d}{d(1/g^2)} \la \bar{\psi} \psi\ra=\frac{g^2m^2}{2}
 \frac{d}{dm}\left( \frac{\la \bar{\psi} \psi\ra}{m} \right).
\ee
 
Now combining the low energy theorems 
(\ref{4},\ref{5}) with (\ref{8}) we obtain the exact relation
between quark and gluon condensates:
\begin{equation}
\label{9}
\frac{1}{2}\frac{d}{dm}\la G_{\mu\nu}^a G^{a\mu\nu}\ra =  m^2 \frac{d}{dm}\left(\frac{\la \bar{\psi} \psi\ra}{m} \right).
\end{equation}
This relation was derived earlier   in \cite{Burkardt2}
 using a  different approach. One can rederive our
previous formula for the gluon condensate (\ref{7}) 
using the exact relation (\ref{9}) and the 
expression (\ref{OPE1})
for the quark condensate in the asymptotic region.

An exact relation between quark and gluon condensates
should not be  considered as a big surprise. Such a 
relation is a direct consequence of the fact that in 
two dimensional $QCD$
the gluons are not dynamical degrees of freedom.

\section{Conclusion}
The main results of this paper are given by Eqs.(\ref{OPE1}),(\ref{7}).
We explicitly calculated the quark and gluon condensates in 
$QCD_2(N)$ coupled to adjoint
matter as a function
of the nontrivial vacuum label $k$.  Our formulae are 
valid only in the vicinity of large $m$,
but the obtained results suggest that all observables do depend on 
the specific vacuum state in which we live. 

In principle, one can further generalize formula
(\ref{OPE}) in order to include higher dimensional 
gluon condensates.
One can derive the next term in the expansion: 
\be 
\label{10}
\la\bar{\psi} \psi\ra  &=& 
   -  \frac{g^2}{\pi^2 m} Tr \left[ T(R) T(R) \right]  -
\frac{1}{12 \pi} \frac{g^2}{ m^3} \la Tr \left[      
G^{\mu\nu} G_{a\mu\nu} \right] \ra    \nonumber \\   
&-&  \frac{1}{12 \pi} \frac{g^2}{ m^5} 
\left( \la Tr \left[ (D_{\alpha}G^{\mu \nu}) ( D^{\alpha} 
G_{\mu \nu} ) \right] \ra + \frac{1}{2} \la Tr 
\left[  G_{\mu \nu} D^2 
G^{\mu \nu} \right] \ra   \right. \\
&-& \left. \frac{1}{3} \la Tr \left[ 
(D_{\alpha} G^{\alpha \mu} ) ( D^{\beta} G_{\beta \mu} ) 
\right]  \ra \right) + \ldots ~~,            \nonumber  
\ee
where $T(R)$ are the generators in the representation R of SU(N).
Note that Eq.(\ref{10}) is valid for arbitrary representation of SU(N).
Unfortunately, we do not know the higher dimensional condensates which
appear in (\ref{10}).
The only thing we know for sure is the fact that the $k$-dependence of 
each condensate should be a periodic function
similar to, but not necessarily the same as, (\ref{OPE1}).
One can reverse this argument. If we knew 
the fermion condensate in $QCD_2^{adjoint}$ for arbitrary $m$ exactly,
(similar to Eq.(\ref{2}) in the 't~Hooft model), we would be able 
to calculate
all high dimensional gluon condensates as a function of $k$.
This problem is probably too complicated
and equivalent to the complete solution of $QCD_2^{adjoint}$
which  presently is  not available.

 We conclude this paper with the following project for future 
investigation.
One may try to use our results
with heavy quark mass to generalize the string
picture originally derived   in \cite{grosst} for the pure gauge
 2D theory without dynamical degrees of freedom.
By introducing a heavy quark we have
essentially introduced the physical degrees of freedom
without noticeably changing the internal gluodynamics.
Therefore, we expect that the introduction of the heavy
quark might be the first step in the direction 
of a string description of dynamical degrees of freedom.
We probably need to implement some dynamics on the 
string-sheet boundary which would correspond to a heavy, 
but dynamical, quark.

\section{Acknowledgements}
This work was supported in part by the Natural Sciences and 
Engineering
Research Council of Canada.

\end{document}